\documentclass[preprintnumbers, prd, showpacs, floatfix,twocolumn,
preprintnumbers, letterpaper, superscriptaddress]{revtex4}
\usepackage{amsfonts}
\usepackage{amsmath}
\usepackage{amssymb,epsf}
\usepackage{latexsym}
\usepackage{graphicx,epsfig}
\usepackage{amssymb}
\usepackage{subfigure}
\usepackage[colorlinks=true,citecolor=blue,linkcolor=blue,urlcolor=black]{hyperref}
\usepackage{epstopdf}

\begin{document}

\title{Novel phase transition in charged dilaton black holes}
\author{Amin Dehyadegari}
\affiliation{Physics Department and Biruni Observatory, Shiraz University, Shiraz 71454,
Iran}
\author{Ahmad Sheykhi}
\email{ashykhi@shirazu.ac.ir}
\affiliation{Physics Department and Biruni Observatory, Shiraz University, Shiraz 71454,
Iran}
\affiliation{Research Institute for Astronomy and Astrophysics of Maragha (RIAAM), P.O.
Box 55134-441, Maragha, Iran}
\author{Afshin Montakhab}
\email{montakhab@shirazu.ac.ir}
\affiliation{Physics Department and Biruni Observatory, Shiraz University, Shiraz 71454,
Iran}

\begin{abstract}
We disclose a novel phase transition in black hole physics by
investigating thermodynamics of charged dilaton black holes in an
extended phase space where the charge of the black hole is
regarded as a fixed quantity. Along with the usual critical
(second-order) as well as the first-order phase transitions in
charged black holes, we find that a finite jump in Gibbs free
energy is generated by dilaton-electromagnetic coupling constant,
$\alpha$, for a certain range of pressure. This novel behavior
indicates a small/large black hole \emph{zeroth-order} phase
transition in which the response functions of black holes
thermodynamics diverge e.g. isothermal compressibility. Such
zeroth-order transition separates the usual critical point and the
standard first-order transition curve. We show that increasing the
dilaton parameter($\alpha$) increases the zeroth-order portion of
the transition curve. Additionally, we find that the second-order
(critical) phase transition exponents are unaffected by the
dilaton parameter, however, the condition of positive critical
temperature puts an upper bound on the dilaton parameter
($\alpha<1$).
\end{abstract}

 \maketitle

\section{Introduction}
Since the discovery of black holes thermodynamics in $1970$'s by
Bekenstein and Hawking \cite{bek,haw}, physicists have been
speculating that there should be some kind of thermodynamic phase
transition in this gravitational system. Hawking and Page were the
first who discovered a first order phase transition of thermal
radiation-large black holes in the background of Schwarzschild
anti-de-Sitter (AdS) spacetime \cite{HP}. Later, by considering
charged black holes, a phase transition was shown to occur between
small-large black holes \cite{Qphi}. Such a phase transition has
been associated to a liquid-gas transition such as the one
occurring in the Van der Waals system \cite{callen,johnston}.
Recently, phase transition of charged AdS black holes has
attracted much attentions. It was shown that thermodynamic
properties of charged AdS black holes admit a first order phase
transition between large and small black holes occurs which is
analogous to the Van der Waals liquid-gas phase transition
\cite{Dolan,mann1}. In this perspective, thermodynamic analysis
are improved in an \textit{extended phase space} in which the
cosmological constant and its conjugate variable are considered as
thermodynamic pressure and volume, respectively. It has been
demonstrated that the first law of black hole thermodynamics is
consistent with the Smarr relation provided the mass of black hole
is identified as the enthalpy \cite{enthalpy}. Phase transition
and critical behaviour of black holes in an extended phase space
have been investigated in ample details (see \cite%
{Dolan2,Ce1,Ur,GB1,GB2,Hendi,Kamrani,Dayyani1,Dayyani2} and references
therein). In all these works (\cite%
{Dolan,mann1,enthalpy,Dolan2,Ce1,Ur,GB1,GB2,Hendi,Kamrani,Dayyani1,Dayyani2}%
) the cosmological constant (pressure) is considered as a variable quantity
and the charge of the black holes is fixed.

In another approach towards thermodynamic phase space of black
holes, it was shown that one can think of variation of charge $Q$
of a black hole and keep the cosmological constant as a fixed
parameter. The motivation for this assumption comes from the fact
that the charge of a black hole is a natural external variable
which can vary \cite{AAA}. Besides, the cosmological constant is
related to the background of AdS geometry and it is more natural
to take it as a constant, rather than a variable quantity
\cite{AAA}. This alternative view of such a phase space and more
physically conventional description of the phase transition
naturally leads to a meaningful response function and a more
accurate analogy with the Van der Waals fluid \cite{AAA}. Indeed,
in this perspective, the critical behavior occurs in
$Q^{2}$-$\Psi$ plane, where $\Psi =1/2r_{+}$ is the conjugate of
$Q^2$ \cite{AAA}. It was shown that a small-large black hole phase transition occurs with an associated critical point $(T_c,Q^2_c, \Psi_c)$ with complete analogy with the Van der Waals fluid system \cite{AAA}.

A discontinuity in the derivatives of Gibbs free energy with
respect to temperature characterizes the type of phase transition
that occurs in thermodynamic system. A first-order phase
transition has a discontinuity in the first derivative which is
entropy, i.e. $(\partial G/\partial T) = S$, and second-order
(critical) phase transition has discontinuity (singularity) of
$(\partial^2 G/ \partial^2 T)$. Therefore, a lesser-known
zeroth-order phase transition has a discontinuity in the Gibbs free
energy itself, which was discovered in superfluidity and
superconductivity \cite{maslov}. The author of Ref.\cite{maslov}
also showed that a zeroth-order phase transition occurs in
Bogolyubov's model of a weakly nonideal Bose gas. On the other
hand, a novel reentrant phase transition has been recently
observed to
accompany the standard first-order phase transitions in black holes\cite%
{BImann,rotmann,highorder,hairyBH,Kubiznak}. Motivated by the
above novel phase transitions, we intend to present a study of
small-large phase transitions in charged dilaton black holes.
Therefore, we analyze the possible phase transitions in extended
phase space for fixed charge where the space-time geometry is
described by Einstein-Maxwell-dilaton gravity \cite{sheykhi}. We
find that the presence of dilaton parameter leads to a region of
the phase diagram which allows for zeroth, first and second-order
phase transition where the zeroth-order separates the first and
second order transition, along the transition curve. The extension
of such a region becomes larger with increasing dilaton parameter.

This Communication is structured as follows: in section \ref{ther}, we
study thermodynamics of $(d+1)$-dimensional charged dilaton black
holes in the presence of Liouville-type dilaton potential. In
section \ref{crit}, we investigate critical behavior of dilaton
black holes. In section \ref{Eq}, we study equation of state of
charged dilaton black holes. The last section is devoted to
summary and conclusions.

\section{Basic thermodynamics of charged dilaton black hole \label{ther}}

The action of ($d+1$)-dimensional spacetime in Einstein-Maxwell theory with
a scalar dilaton field ($\varphi $), reads \cite{manndilaton}%
\begin{eqnarray}
\mathcal{I} &=&\frac{1}{16\pi }\int d^{d+1}x\sqrt{-g}\Big(\mathcal{R}-\frac{4%
}{d-1}(\nabla \varphi )^{2}-\mathcal{V} (\varphi )  \notag \\
&&-e^{-4\alpha \varphi /(d-1)}F_{\mu \nu }F^{\mu \nu }\Big),  \label{ac}
\end{eqnarray}%
where $F_{\mu \nu }=\partial _{\lbrack \mu }A_{\nu ]}$, $A_{\nu }$ is the
vector potential and $\alpha $ is the coupling parameter of dilaton with
Maxwell field. Hereon, $\mathcal{V}(\varphi )$ is the dilaton potential
which has the following form \cite{manndilaton,sheykhi}%
\begin{equation}
\mathcal{V}(\varphi )=2\Lambda e^{4\alpha \varphi /(d-1)}+\frac{%
(d-1)(d-2)\alpha ^{2}}{b^{2}(\alpha ^{2}-1)}e^{4\varphi /[(d-1)\alpha ]},
\end{equation}%
where  $b$  is a positive arbitrary constant. In the absence of
dialton field ($\alpha =0$) the above potential reduces to
$\mathcal{V}(\varphi ) \rightarrow2\Lambda $, and thus one may
interpret $\Lambda $ as the cosmological constant. The
($d+1$)-dimensional spherical symmetric metric is given by
\begin{equation}
ds^{2}=-f(r)dt^{2}+\frac{dr^{2}}{f(r)}+r^{2}R(r)^{2}d\Omega _{d-1}^{2},
\end{equation}%
where $d\Omega _{d-1}^{2}$ is the line element of an unit
($d-1$)-sphere with
the volume $\omega _{d-1}$. Applying ansatz $R(r)=e^{2\alpha \varphi /(d-1)}$%
, one can show that \cite{sheykhi}%
\begin{eqnarray}
f(r) &=&\frac{2\Lambda \left( \alpha ^{2}+1\right) ^{2}r^{2(1-\gamma )}}{%
(d-1)\left( \alpha ^{2}-d\right) b^{-2\gamma }}-\frac{\left( d-2\right)
\left( \alpha ^{2}+1\right) ^{2}{b}^{-2\gamma }{r}^{2\gamma }}{\left( \alpha
^{2}-1\right) \left( \alpha ^{2}+d-2\right) }  \notag \\
&&-\frac{m}{r^{\left( d-1\right) \left( 1-\gamma \right) -1}}+\frac{%
2q^{2}(\alpha ^{2}+1)^{2}r^{2(d-2)(\gamma -1)}}{(d-1)(\alpha
^{2}+d-2)b^{2\gamma (d-2)}},
\end{eqnarray}%
\begin{equation}
\varphi (r)=\frac{(d-1)\alpha }{2(1+\alpha ^{2})}\ln \left(\frac{b}{r}%
\right),\text{ \ \ \ \ }A_{t}=\frac{qb^{\gamma (3-d)}}{\Pi r^{\Pi }},
\end{equation}%
where $\gamma =\alpha ^{2}/(\alpha ^{2}+1)$, $\Pi =(d-3)(1-\gamma )+1$, $b$
is a positive arbitrary constant, $m$ and $q$, respectively, are related to
the total mass and electric charge of the black hole \cite{sheykhi}%
\begin{equation}
M=\frac{b^{\gamma (d-1)}(d-1)\omega _{d-1}}{16\pi \left( \alpha
^{2}+1\right) }m,\text{ \ \ }Q=\frac{q\omega _{d-1}}{4\pi }.
\end{equation}%
Inasmuch as the event horizon is defined by the largest root of $f(r_{+})=0$%
, one can write $m$ in terms of $r_{+}$. Temperature, entropy and electric
potential of Einstein-Maxwell-dilaton black holes are obtained as \cite%
{sheykhi}
\begin{eqnarray}
T &=&-\frac{(d-2)\left( \alpha ^{2}+1\right) b^{-2\gamma }}{4\pi \left(
\alpha ^{2}-1\right) }r_{+}^{2\gamma -1}-\frac{\Lambda \left( \alpha
^{2}+1\right) b^{2\gamma }}{2\pi (d-1)}r_{+}^{1-2\gamma }  \notag \\
&&-\frac{q^{2}\left( \alpha ^{2}+1\right) b^{-2\gamma (d-2)}}{2\pi (d-1)}%
r_{+}^{(2d-3)(\gamma -1)-\gamma },  \label{Tem}
\end{eqnarray}%
\begin{equation}
S=\frac{b^{\gamma (d-1)}r_{+}^{(d-1)(1-\gamma )}\omega _{d-1}}{4},
\label{entropy}
\end{equation}%
\begin{equation}
U=\frac{qb^{\gamma (3-d)}}{\Pi {r_{+}}^{\Pi }}.
\end{equation}%
Using Eq.(\ref{entropy}) and relation $V=\int 4Sdr_{+}$, one obtains the
volume as
\begin{equation}
{V}=\frac{(1+\alpha ^{2})b^{(d-1)\gamma }\omega _{d-1}}{d+\alpha ^{2}}%
r_{+}^{(d+\alpha ^{2})/(1+\alpha ^{2})}.
\end{equation}
In the extended phase space, the mass of black hole is considered as
enthalpy \cite{enthalpy} and hence the first law of thermodynamics and Smarr
formula take the form
\begin{eqnarray}
dM &=&TdS+UdQ+VdP, \\
M &=&\frac{(d-1)(1-\gamma )}{\Pi }TS+UQ+\frac{(4\gamma -2)}{\Pi }VP,
\end{eqnarray}%
in which $P$ is the thermodynamic pressure, given by
\begin{equation}
P=-\frac{\left( d+\alpha ^{2}\right) b^{2\gamma }}{8\pi \left( d-\alpha
^{2}\right) r_{+}^{2\gamma }}\Lambda .
\end{equation}%
In the absence of dilaton ($\alpha =0$), the above pressure becomes $%
P=-\Lambda /8\pi $, which is the pressure of the
Reissner-Nordstrum-AdS (RN-AdS) black hole  \cite{mann1}.

\section{Instability and phase transition in dilaton black hole}

\label{crit} The sign of the response functions must be positive
for local stability of  a thermodynamic system \cite{callen}.
Since we consider an extended phase space, it is important to
study the behavior of the
isothermal compressibility%
\begin{equation}
\kappa _{{}_{T}}=-\frac{1}{V}\left. \frac{\partial V}{\partial P}\right\vert
_{T}.
\end{equation}%
Negative sign of $\kappa _{{}_{T}}$ indicates local thermodynamic
instability and therefore a phase transition in the system. To see
the exact behavior of thermodynamic system with regard to local
instability, we need to calculate the Gibbs free energy
$G=G(T,P)$, which is obtained as
\begin{eqnarray}
G &=&M-TS=\Bigg\{\frac{(d-2)(1+\alpha ^{2})b^{\gamma (n-3)}}{16\pi (\alpha
^{2}+d-2)r_{+}^{(n-3)(\gamma -1)-1}}  \notag \\
&&+\frac{P(\alpha ^{4}-1)b^{\gamma (d-1)}}{(d-1)(d+\alpha
^{2})r_{+}^{d(\gamma -1)-\gamma }}  \notag \\
&&+\frac{q^{2}(2d-3+\alpha ^{2})(\alpha ^{2}+1)b^{\gamma (3-d)}}{8\pi
(d-2+\alpha ^{2})(d-1)r_{+}^{(n-3)(1-\gamma )+1}}\Bigg\}\omega _{n-1},
\notag \\
&&
\end{eqnarray}%
\begin{figure}[t]
\epsfxsize=8.5cm \centerline{\epsffile{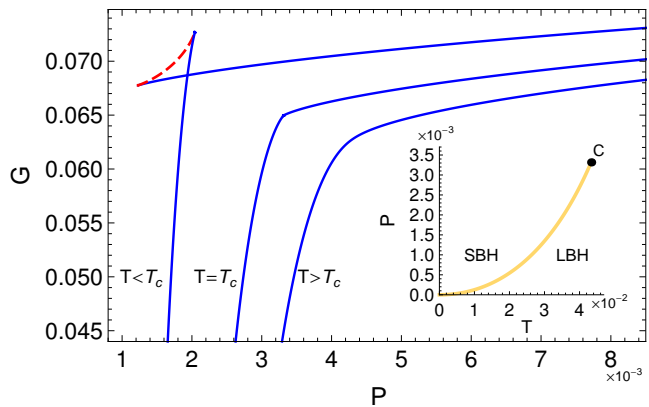}}
\caption{Gibbs free energy as a function of pressure for $\protect\alpha =0,$
$q=1$ and $d=3$ and different values of temperature. Below the critical
temperature ($T_{c}$), it shows multi-valued behavior indicating a first
order phase transition (SBH/LBH). The positive (negative) sign of $\protect%
\kappa _{{}_{T}}$ is identified by blue solid (dashed red) line. Note that
various curves are shifted for clarity. The corresponding phase diagram ($P$-%
$T$) is shown in the inset.}
\label{fig1}
\end{figure}
Let us first consider the Gibbs free energy in the absence of dilaton ($%
\alpha =0$), depicted in Fig. \ref{fig1} for $d=3$. According to Fig.(\ref%
{fig1}), the Gibbs energy is single-valued and increases
monotonically with the increasing pressure for $T>T_{c}$ and is
locally stable ($\kappa
_{{}_{T}}>0$) everywhere as indicated by the solid blue line. However, for $%
T<T_{c}$, it becomes multi-valued with negative $\kappa _{{}_{T}}$
shown by dashed red line. Since the minimum value of $G$ is chosen
by the system at equilibrium, this indicates a first order phase
transition which occurs between small black hole (SBH) and large
black hole (LBH) where the slope of $G$ is discontinuous at the
transition point. At $T=T_{c}$, a second-order phase transition
occurs between SBH and LBH where $G$ is single-valued and
continuous, but is non-analytic. The corresponding phase ($P$-$T$)
diagram for RN-AdS is illustrated in the inset of Fig.\ref{fig1}.
SBH is distinguished from LBH by a transition line with a critical
point at the end of the transition curve. Note that this
transition curve looks analogous to the Van der Waals $P$-$T$
diagram. The qualitative behavior for higher dimensional black
holes is the same as $d=3$.

\begin{figure}[t]
\epsfxsize=8.5cm \centerline{\epsffile{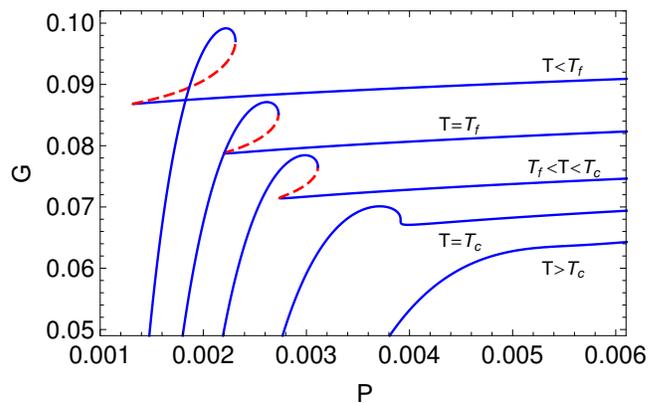}}
\caption{Gibbs free energy as a function of pressure for different values of
temperature and $\protect\alpha =0.4$, $b=1,$ $q=1$ and $d=3$. The positive
(negative) sign of $\protect\kappa _{{}_{T}}$ is identified by blue solid
(dashed red) line. For $T\leq T_{c}$, $G(P)$ develops non-analytic behavior
in the different ways, at $T=T_{c}$, for $T_{\mathrm{f}}<T<T_{c}$ and
finally for $T<T_{\mathrm{f}}$. Note that various curves are shifted for
clarity.}
\label{fig2}
\end{figure}

Now, we turn to examine the effects of the dilaton field parameter
($\alpha \neq 0$) on the phase transition of black holes in
extended phase space. For this purpose, we plot the Gibbs free
energy as a function of pressure for $\alpha =0.4$ and $d=3$ in
Fig. \ref{fig2}. One can see that for $T\geq T_{c}$, the behavior
is similar to the previous case ($\alpha =0$), namely the Gibbs
free energy is single-valued and increases monotonically with
increasing the pressure. Here $T_{c}$ is the critical point where
a second order phase transition occurs. In the range of
temperature $T\leq T_{\mathrm{f}}$,
where the loop is formed at $T=T_{\mathrm{f}}$, a \textit{%
first order} phase transition occurs similarly to the Van der
Waals fluid system. However, for the temperature range $T_{\rm
f}<T<T_{c}$ an interesting phenomenon occurs where a finite jump
in $G$ leads to a zeroth-order phase transition. Such zeroth-order
phase transition has previously been considered in the theory of
superfluidity and superconductivity \cite{maslov}, and more
recently has been reported as a part of reentrant phase transition
in black holes\cite{BImann,rotmann,highorder,hairyBH,Kubiznak}.
\begin{figure}[t]
\epsfxsize=8.5cm \centerline{\epsffile{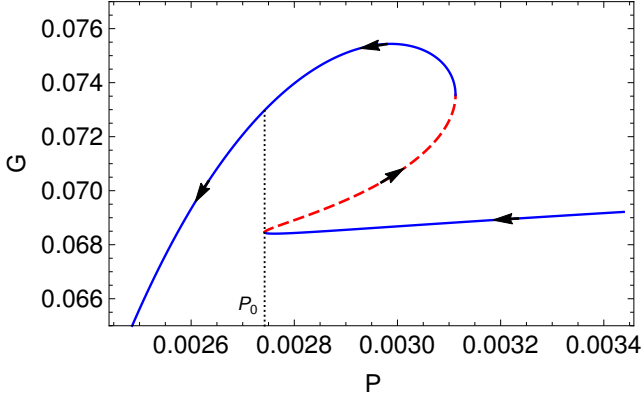}}
\caption{A close-up of $T\approx 0.064$ $\in \left[ T_{\mathrm{f}},T_{c}%
\right] $ in Fig. \protect\ref{fig2} displays a zeroth order phase
transition which is accompanied by a finite jump in $G$ at $P_{0}\approx
0.00274$. The positive (negative) sign of $\protect\kappa _{{}_{T}}$ is
identified by blue solid (dashed red) line and the arrows show the direction
of the increasing $r_{+}$.}
\label{fig3}
\end{figure}
\begin{figure}[t]
\epsfxsize=8.5cm \centerline{\epsffile{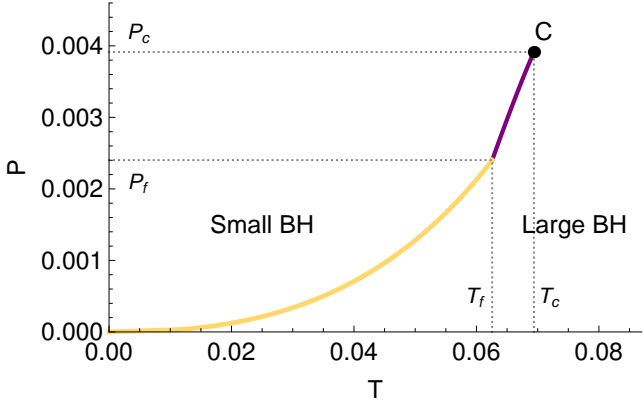}}
\caption{Phase diagram corresponding to Fig. \protect\ref{fig2} for
transition between SBH and LBH. Three different transition are identified:
first order (gold curve), zeroth order (purple curve) and a second order
critical point at ($T_{c},P_{c}$).}
\label{fig4}
\end{figure}

To be more specific, let us consider the case of $T\approx 0.064$
$\in \left[ T_{\rm f},T_{c}\right] $ in Fig. \ref{fig3}.
Decreasing the pressure of the
system (despite its multi-valuedness) follows a locally stable regime ($%
\kappa _{{}_{T}}>0$) by choosing the lower $G$ value. However, at
$P=P_{0}$ a jump to a higher (stable) value of $G$ is required
thus leading to a discontinuous $G$ and a zeroth order phase
transition. Note that this type of phase transitions is absent in
normal liquid-gas transitions such as the Van der Waals fluid.

The phase diagram of charged dilaton black hole is shown in Fig. \ref{fig4}.
A Van der Waals like first order phase transition occurs for $\left(
T,P\right) <\left( T_{\mathrm{f}},T_{\mathrm{f}}\right) $. This is followed
by zeroth order phase transition curve with finite jump in $G$ for the range
$\left( T_{\mathrm{f}},T_{\mathrm{f}}\right) <\left( T,P\right) <\left(
T_{c},T_{c}\right) $. Interestingly, the zeroth-order transition curve
terminates at the critical point $\left( T_{c},T_{c}\right) $.

\section{Equation of state}

\label{Eq} 
\begin{figure}[t]
\epsfxsize=8.5cm \centerline{\epsffile{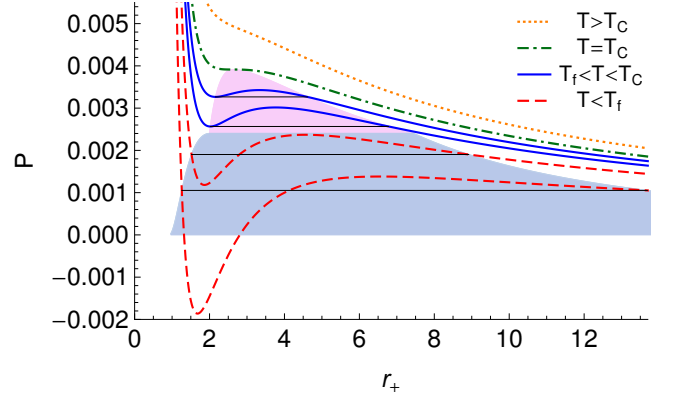}}
\caption{$P$-$r_{+}$ diagram of charged dilaton black holes for various
temperatures and $\protect\alpha =0.4$, $b=1,$ $q=1$ and $d=3$. The regions
of zeroth and first order phase transition are characterized by different
colors. The isobars (black thin line) remedy the unphysical locally and
globally unstable regimes, see Fig. \protect\ref{fig2}.}
\label{fig5}
\end{figure}

The equation of state $P=P(T,r_{+})$ (for fixed $q$ ) can be
obtained by using Eq.(\ref{Tem}) as%
\begin{eqnarray}
P &=&\frac{(d+\alpha ^{2})(d-1)T}{4(d-\alpha ^{2})(1+\alpha ^{2})r_{+}}+%
\frac{(d-2)(d-1)(d+\alpha ^{2})b^{-2\gamma }}{16\pi (\alpha ^{2}-1)(d-\alpha
^{2})r_{+}^{2-2\gamma }}  \notag \\
&&+\frac{(d+\alpha ^{2})q^{2}b^{-2\gamma (d-2)}}{8\pi (d-\alpha
^{2})r_{+}^{1-(2d-3)(\gamma -1)+\gamma }}.  \label{Eqstate}
\end{eqnarray}%
The $P-r_{+}$ isothermal diagrams for $\alpha =0.4$ and $d=3$ are shown in
Fig. \ref{fig5}. The critical point is essentially an inflection point where
($\partial P/\partial r_{+}=0$ and $\partial ^{2}P/\partial r_{+}^{2}=0$),
and can be obtained as \cite{Kamrani}%
\begin{eqnarray}
P_{c} &=&\left[ \frac{(d-1)(d-2)}{2(2d-3+\alpha ^{2})(d-1+\alpha ^{2})}%
\right] ^{[\gamma -(2d-3)(\gamma -1)+1]/2\Pi }  \notag \\
&&\times \left[ \frac{(d+\alpha ^{2})(2d-3+\alpha ^{2})(d-2+\alpha ^{2})}{%
8\pi (1+\alpha ^{2})(d-\alpha ^{2})b^{2\gamma /\Pi }q^{2(1-\gamma )/\Pi }}%
\right] , \\
T_{c} &=&\left[ \frac{(d-1)}{2q^{2}(d-1+\alpha ^{2})}\right] ^{(1-2\gamma
)/2\Pi }\left( \frac{(\alpha ^{2}+d-2)}{\pi (1-\alpha ^{2})b^{\gamma
(n-1)/\Pi }}\right)  \notag \\
&&\times \left[ \frac{(2d-3+\alpha ^{2})}{(d-2)}\right] ^{[(2d-3)(\gamma
-1)-\gamma ]/2\Pi },  \label{ctem} \\
r_{+c} &=&\left[ \frac{2q^{2}(d-1+\alpha ^{2})(2d-3+\alpha ^{2})b^{6\gamma
-2d\gamma }}{(d-2)(d-1)}\right] ^{1/2\Pi }.
\end{eqnarray}%
According to Eq. (\ref{ctem}), critical temperature has positive values if
dilaton parameter is restricted to $\alpha <1$. One can calculate the
critical exponents $\alpha ^{\prime }=0$, $\beta ^{\prime }=1/2$, $\delta
^{\prime }=3$ and $\gamma ^{\prime }=1$ associated with second order
transition which are mean-field values such as the Van der Waals fluid
system \cite{Kamrani}.

\begin{figure}[t]
\epsfxsize=8.5cm \centerline{\epsffile{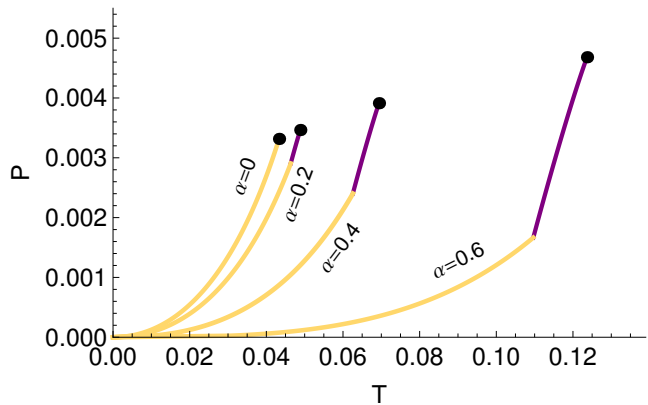}}
\caption{SBH/LBH phase diagram for various values of dilaton parameter $%
\protect\alpha $ and $b=1,$ $q=1$ and $d=3$. Increasing $\protect\alpha $
leads to a larger portion of the transition curve belonging to zeroth order
phase transition. The critical points are highlighted by a black solid
circle.}
\label{fig6}
\end{figure}
For $T<T_{c}$, we observe local instability ($\kappa _{{}_{T}}<0$)
and negative pressure in some range of quantities for the system.
Such an unphysical behavior is remedied by choosing the globally
stable Gibbs free energy as illustrated in Fig. \ref{fig2}. For
$T<T_{\mathrm{f}}$ where a first order phase transition occurs, a
globally unphysical part is replaced by the isobar line for which
$G_{\mathrm{SBH}}=G_{\mathrm{LBH}}$. Following the behavior of
Gibbs free energy, we observe that the modified isobar line for
the zeroth order phase transition starts at the point of local
minimum of isotherm for range $T_{\mathrm{f}}<T<T_{c}$. One can
see from Fig. \ref{fig5} that $\kappa _{T}$ diverges ($\partial
P/\partial r_{+}=0$) in the case of zeroth order phase transition.
Similarly, one can plot the entropy ($S$) versus temperature ($T$)
as well, and will subsequently see that the heat capacity at the
constant pressure diverges in zeroth order phase transition.

Fig. \ref{fig6} shows SBH/LBH phase diagram for various values of
dilaton parameter $\alpha $. As is seen from the figure increasing
$\alpha $ from zero (RN-AdS BH) will lead to creation and
elongation of the zeroth order phase transition as the critical
point moves higher and the first order transition curve (purple)
bends lower and further in the $P$-$T$ space. Finally, it is worth
mentioning that similar  qualitative behavior for the higher
dimensional ($d>3$) charged dilaton black holes can be observed.


\section{Summary\label{Con}}

In this paper, we have investigated the thermodynamic phase behaviour of
charged dilaton black holes in the presence of Liouville-type dilaton
potentials. Due to the presence of the dilaton field, these solutions are
neither asymptotically flat nor (A)dS \cite{sheykhi}. We have disclosed the
effects of the dilaton field on the phase transition properties of charged black holes
in an extended phase space. In addition to the usual small/large black holes
transition (first and second order) \cite{Dolan,mann1}, we observed a
\textit{zeroth order} phase transition between small/large black hole in
which isothermal compressibility and heat capacity at the constant pressure
diverge. Besides, the Gibbs free energy has a finite jump, at the point
where a \textit{zeroth order} phase transition occurs. We also obtained that
a zeroth order phase transition emerges in a longer portion of the
transition line by increasing a coupling constant of dilaton field. It is
worth noting that a coupling constant is restricted to $\alpha <1$, for
which the critical temperature is positive. Finally, we showed a set of
critical exponents which are the same with Van der Waals fluid system.

\begin{acknowledgments}
Shiraz University research council is kindly acknowledged. The work of AS
has been supported financially by Research Institute for Astronomy \&
Astrophysics of Maragha (RIAAM), Iran.
\end{acknowledgments}

\end{document}